

\input panda
%
%
\loadamsmath\chapterfont{\bfone} \sectionfont{\scaps}
\def\sssty{\scriptscriptstyle}   \def\v{{\cal V}}
\nopagenumbers
{\baselineskip=12pt
\line{\hfill PUPT-1313}
\line{\hfill hepth@xxx/9204052}
\line{\hfill April, 1992}}
{\baselineskip=14pt
\ifdoublepage \bjump\bjump\else\bjump\bjump\jump\fi
\centerline{\capsone CORRELATION FUNCTIONS FROM TWO-DIMENSIONAL}
\sjump
\centerline{\capsone STRING WARD IDENTITIES }
\bjump\bjump
\centerline{\scaps Igor R. Klebanov and Andrea Pasquinucci}
\sjump
\centerline{\sl Joseph Henry Laboratories, Department of Physics,}
\centerline{\sl Princeton University, Princeton, NJ 08544, USA}
\vfill
\ifdoublepage \eject\null\vfill\bjump\fi
\centerline{\capsone ABSTRACT}
\sjump \noindent We rederive the $w_\infty$ Ward identities,
starting from the existence of trivial linearized gauge invariances,
and using the method of canceled propagators in the operator formalism.
Recursion relations for certain classes of correlation functions
are derived, and these correlation function are calculated exactly.
We clarify the relation of these results with another derivation of the
Ward identities, which relies directly on charge conservation.
We also emphasize the importance of the kinematics of canceled
propagators in ensuring that the Ward identities are non-trivial.
Finally, we sketch an extension of Ward identities to open strings.

\jump
\ifdoublepage\vfill\else\bjump\jump\fi
\pageno=0 \eject }
\yespagenumbers\pageno=1
\introduction
Recently there has been considerable interest in a model of bosonic
strings propagating in two-dimensional space-time. This model arises
upon the third quantization of two-dimensional quantum gravity
coupled to one massless scalar field. The model has been
solved exactly in the discretized approach implemented through
large--$N$ matrix quantum mechanics [\Ref{GMil},\Ref{GKleb}]. Now
it seems important to understand this exact solubility in the more
conventional continuum approach to the sum over surfaces --- the
Liouville gravity [\Ref{Liouv}]. In the conformal gauge the theory
reduces to a sigma model which describes strings propagating in two
dimensions [\Ref{ddk}]. One of the coordinates is the
original massless scalar, while the other, $\phi$, arises from the
conformal factor of the world sheet metric. There is a dilaton
background linear in $\phi$ which breaks translation invariance.
Thus, the two coordinates enter on a different footing and
the model is not Lorentz invariant. However, as explained in refs.
[\Ref{W},\Ref{KP}], the model possesses an infinite symmetry, which
is isomorphic to the area-preserving diffeomorphisms (a wedge
subalgebra of $w_\infty$). The symmetry charges, which are
constructed from the discrete states
[\Ref{discrete},\Ref{GKN},\Ref{lz},\Ref{WZ}], are expected to
provide an explanation for the exact solubility of the model.

The symmetry structure found in the continuum approach strongly
resembles the symmetries present in the matrix model solution
[\Ref{aj}]. A complete equivalence is still missing because in the
matrix model the Liouville potential (``tachyon background'') is
turned on, while the continuum treatment has mostly dealt with the
free field theory where the potential has been set to zero. The free
field approach is sufficient to study the so-called bulk amplitudes
[\Ref{Sasha},\Ref{GK}], which obey an appropriate energy sum rule.
The bulk amplitudes are believed to determine indirectly all the
correlation functions of the theory [\Ref{Goulian},\Ref{kdf}].
Attempts to study the effects of the Liouville potential on the
symmetry structure have not yet produced conclusive results
[\Ref{pert}]. In this paper we will restrict ourselves to the
bulk amplitudes and to the free field theory where the
potential is turned off, and the $w_\infty$ symmetry structure is
completely transparent.

After the discovery of the infinite number of conserved charges, a
natural step is to derive the Ward identities based on maintaining the
charge conservation inside correlation functions.\note{For another
approach to constraining the correlation functions, see refs.
[\Ref{KMS},\Ref{Kachru}].} In ref. [\Ref{K}] various
sources of local charge non-conservation on the world sheet were
examined, and it was shown that the requirement of global charge
conservation leads to recursion relations among correlation functions.
These recursion relations were used to calculate the bulk tachyon
amplitudes in a few simple steps. In other words, the $w_\infty$
symmetry Ward identities allow one to calculate a class of
formidable multiple integrals, which are generalizations of
the Dotsenko-Fateev integrals [\Ref{Dots}], with hardly any work.
The existence of the recursion relations is due to the fact that
the $w_\infty$ charges generally act on the massless ``tachyons''
non-linearly, i.e. a charge acting on a state with $n$ particles
produces a state with $m\leq n$ particles. In refs.
[\Ref{WZ},\Ref{ver}] the Ward identities were derived, and their
non-linear structure was explained, in a somewhat different way.
Each tree level Ward identity was understood as the vanishing of a
sum of products of correlation functions which arise when a sphere is
pinched into two spheres in all possible ways.
In ref. [\Ref{ver}] the Ward identity was identified with the master
equation of Batalin-Vilkovisky quantization [\Ref{bv}].

The aim of this paper is to elucidate the relation between the methods
of refs. [\Ref{WZ},\Ref{K},\Ref{ver}], and to speculate on the general
circumstances under which Ward identities provide non-trivial
constraints on correlation functions. We also sketch extension to
open string theories.

Perhaps, the simplest statement of the Ward identities is through the
observation [\Ref{W},\Ref{Polyakov}] that the theory possesses
states that are both pure gauge and identically zero. The vertex
operator for such a state is
$$\{Q_{BRST}, cW_{J, m}\}=0\nfr{uno}
and formally carries the physical ghost number $n_{gh}=2$.
It vanishes identically because the zero-picture current $cW_{J, m}$
is BRST invariant. A special feature of the theory under consideration
is the presence of an infinite number of such BRST invariant
zero-picture currents of ghost number 1. In general, linearized closed
string gauge invariance assumes the form
$$\delta \Psi=\{Q_{BRST}, \lambda\}
\nfr{gauge}
where the ghost numbers of $\Psi$ and $\lambda$ are 2 and 1
respectively. Therefore, there is an infinite number of gauge
parameters, $\lambda=cW_{J, m}$, for which the linearized gauge
invariance is trivial. In fact, as stated in refs.
[\Ref{W},\Ref{Polyakov}], the presence of such trivial gauge
transformations guarantees that there are discrete states of ghost
number 2 which cannot be gauged away. While at a general momentum
there are enough gauge invariances to gauge away all the oscillator
states, at the discrete momenta carried by $cW_{J, m}$ some gauge
invariances become trivial, eq. \uno, and there appear physical
oscillator states that cannot be gauged away. Thus, in a theory
with generally continuous momenta, the presence of symmetry charges
seems intrinsically connected with the presence of discrete states,
which are physical only at special discrete momenta.
In higher-dimensional theories the only known cases of
this phenomenon occur at zero momentum.
Some familar examples are the conservation of charge and the
associated extra photon state at zero momentum, and the conservation
of momentum and the extra graviton-dilaton states at zero momentum.

The Ward identities follow after inserting eq. \uno\  into
correlation functions [\Ref{WZ},\Ref{ver}]. As usual, the BRST
anti-commutator can be re-written as a sum over the boundaries of
the moduli space of a sphere with $n$ punctures where the sphere
is pinched into two spheres with $m$ and $n-m$ punctures respectively.
In the literature on string theory these boundaries of moduli space
are sometimes referred to as the ``canceled propagators''. The reason
for this terminology is most apparent in the operator formalism which
we will review below. In the 26-dimensional string theory, the
``canceled propagator argument'' is the statement that such boundary
terms on moduli space typically vanish (at least in the context of
an appropriate analytic continuation), because the momentum that
flows through the pinch is off-shell. In the 2-dimensional string
theory the situation is very different. As emphasized in ref.
[\Ref{K}], there is an infinite set of special kinematical
arrangements where the momentum that flows through the pinch is
precisely on-shell. When this is the case, the ``canceled propagator
contribution'' is non-vanishing and explicitly calculable.
The Ward identity is nothing but the statement that the sum of all the
canceled propagator contributions, which arise after the insertion of
eq. \uno\ into a correlation function, vanishes. We emphasize
that one can search for the non-trivial contributions to the Ward
identity simply on the basis of studying the kinematics:
whenever the momentum that flows through the canceled propagator
corresponds to an on-shell state, one expects, and usually finds, a
non-vanishing contribution.
\ifx\risposta\standardrisposta\newpage\fi
\chapter{Closed string Ward identities: derivation in the operator
formalism}
Before we proceed to explicit calculations, let us state
our conventions. Following ref. [\Ref{K}] we set $\alpha^\prime=4$
so that $\vev{X(z,\bar{z}) X(w,\bar{w})}=-2 \log \vert z-w\vert^2$
where $X$ is the $c=1$ matter field and $\phi$ is the Liouville
field. In these conventions the tachyon operators are given by
$$\eqalign{& T^\pm_k(z,\bar{z})\ =
\ e^{ikX+\epsilon_\pm\phi}(z,\bar{z})\cr
&\epsilon_\pm=-1\pm k\cr}
\efr
and we define $V^\pm_k(z,\bar{z})\ \buildchar{=}{\rm def}{ }\
c(z)\bar{c}(\bar{z}) T^\pm_k(z,\bar{z})$. We are interested in
studying the bulk correlation functions of tachyons on a sphere,
$$
\vev{V^\pm_{k_1} \cdots V^\pm_{k_n} }\ .
\efr
They satisfy the following conservation equations
$$
\sum_{i=1}^n k_i\ =\ 0 \qquad\qquad \sum_{i=1}^n \epsilon_i\ =\ -2\ .
\nfr{sumrules}
We will write the correlation functions more explicitly in
two different ways. \newline
In the path integral formalism,
$$
\vev{V^\pm_{k_1} \cdots V^\pm_{k_n} }\ =\ \vev{ V^\pm_{k_1} (+\infty)
V^\pm_{k_2}(1) \int T^\pm_{k_3} \dots \int T^\pm_{k_{n-1}}\,
V^\pm_{k_n}(0) }
\nfr{anove}
where $\int T^\pm_k\ \buildchar{=}{\rm def}{ }\
\int d^2 y \, T^\pm_k(y,\bar{y})$. \newline
In the operator formalism,\note{See for example refs.
[\Ref{GSW},\Ref{divec}] and references therein.}
$$\eqalignno{
\vev{V^\pm_{k_1} \dots V^\pm_{k_n} }\ =\ &\bra{V^\pm_{k_1} }
\, V^\pm_{k_2}(1)\, \Delta\,  V^\pm_{k_3}(1)\, \Delta \cdots
\Delta\, V^\pm_{k_{n-1}}(1)\, \ket{V^\pm_{k_n}} \cr
& +\ {\rm permutations}&\nameali{adieci}\cr}
$$
where $\ket{V^\pm_{k_n}}\ \buildchar{=}{\rm def}{ }\
\lim_{z\rightarrow 0} V^\pm_{k_n}(z) \,\ket{0}$, $\bra{V^\pm_{k_1} }\
\buildchar{=}{\rm def}{ }\  \lim_{z\rightarrow\infty} \bra{0}\,
V^\pm_{k_1}(z)$. The propagator, including the
ghosts' contribution, is
$$\Delta\ =\ {b_0^+ b_0^- \over L_0+\bar L_0}\,\Pi_{L_0, \bar L_0}
\efr
where $b_0^\pm \ \buildchar{=}{\rm def}{ }\ b_0 \pm \bar{b}_0,$
and the Virasoro generators include the ghost pieces.
$\Pi_{L_0, \bar L_0}$ is the projector onto states that satisfy
$L_0=\bar L_0$.

In ref. [\Ref{W}] it was shown that in the $c=1$ closed
string theory there exists an infinite number of conserved currents.
The corresponding zero-picture (``fixed'') vertex operators are
$$
\Omega^{(0)}_{J,m}(z,\bar{z})\ =\ c(z) W_{J,m}(z,\bar{z})\ ,
\qquad\quad W_{J,m}(z,\bar{z})\ =\ \Psi_{J+1,m}(z)
\overline{\cal O}_{J,m}(\bar{z})
\nfr{adue}
and $\overline\Omega^{(0)}_{J,m}$ which is built analogously.
$\Psi_{J+1,m}(z)$ are the gravitationally dressed primary fields
of the $c=1$ matter system whereas the $\overline{\cal O}_{J,m}
(\bar{z})$ form the ``{\sl Ground Ring\/}" [\Ref{W}].
The current algebra is [\Ref{W},\Ref{KP}]
$$
W_{J_1,m_1}(z,\bar{z})W_{J_2,m_2}(w,\bar{w})\ =\ {\scriptstyle
2((J_1+1)m_2 - (J_2+1) m_1) \over z-w}\, W_{J_1+J_2, m_1+m_2}
(w,\bar{w})\ .
\efr
As shown in ref. [\Ref{WZ}], the associated charges are given
by~\note{We follow the conventions of ref. [\Ref{WZ}], section 4,
for the orientation of the contour integrals.}
$$
{\cal A}_{J,m}\ =\ \oint {dz\over 2\pi i}\, \Omega^z_{J,m}(z,\bar{z})
\ -\ \oint {d\bar{z}\over 2\pi i}\,
\Omega^{\bar z}_{J,m}(z,\bar{z})
\efr
where $\Omega^z_{J,m}(z,\bar{z})= W_{J,m}(z,\bar{z})$,
and $\Omega^{\bar z}_{J,m}(z,\bar{z})=c(z) \Psi_{J+1,m}(z)
\overline{X}_{J,m}(\bar{z})$, where $\overline{X}_{J,m}(\bar{z}) =
\bar{b}_{-1}\overline{\cal O}_{J,m}(\bar{z})$.
The algebra of the charges is
$$
[{\cal A}_{J_1,m_1}{\cal A}_{J_2,m_2}]\ =2\bigl(
(J_1+1)m_2 - (J_2+1) m_1\bigr) {\cal A}_{J_1+J_2, m_1+m_2}\ .
\efr

Following refs. [\Ref{WZ},\Ref{ver}], we will generate the Ward
identities by inserting eq. \uno\ into correlation functions.
Then the Ward identities for tachyon correlation functions assume the
form
$$
\vev{\{Q_{BRST} , cW_{J, m}\}V^\pm_{k_1} \cdots V^\pm_{k_n} }\ =\ 0\ .
\efr
More explicitly, in the operator formalism we have
$$\eqalignno{
0\ =\ &\bra{V^\pm_{k_1} } \{Q_{BRST} , cW_{J, m}\} (1)\, \Delta \,
V^\pm_{k_2}(1)\,\Delta \cdots  \Delta\, V^\pm_{k_{n-1}}(1)\,
\ket{V^\pm_{k_n}} \cr
&\ + \ {\rm permutations}\ .&\nameali{auno}\cr}
$$
Having written the Ward identity in this form, we can now apply the
``canceled propagator argument". As already mentioned in the
introduction, on general grounds a BRST anti-commutator can be
rewritten as a sum over the boundaries of the moduli space of a sphere
with $n$ punctures where the sphere is pinched into two spheres
with $p$ and $n-p$ punctures respectively. In the operator
formalism this is obtained explicitly by commuting $Q_{BRST}$ through
each of the propagators $\Delta$. First observe that
$$
\left[ Q_{BRST} , \Delta \right] \ =\ \Pi_{L_0, \bar L_0}\ b_0^- \
=\ \sum_i\ket{\Phi_i}\bra{\Phi^i}\, b_0^-
\nfr{bzero}
where the sum is over a complete set of states that satisfy
$(L_0-\bar L_0)\ket{\Phi_i}=0$.
Thus, $Q_{BRST}$ can literally cancel a propagator, replacing it with
an insertion of $b_0^-$. The conjugate states
$\ket{\Phi_i}$ and $\ket{\Phi^i}$ satisfy $\bivev{\Phi^j}{\Phi_i}
= \delta^j_i$ (for continuous spectrum the Kroenecker symbol is
replaced by the Dirac delta function).
Each pair of conjugate states has their ghost numbers
add up to 6, their momenta $k$ add up to 0, and their energies
$\epsilon$ add up to $-2$. For instance, the state conjugate to
$\ket{V^\pm_k}$ is
$$\ket{\v^\pm_{-k}}\ =\ c \partial c\,\bar c \bar \partial\bar c
\, T^\pm_{-k}(0) \ket{0}\ .
\efr
It is also convenient to define the states $\ket{\widetilde{V}^\pm_k}
\ \buildchar{=}{\rm def}{ }\ b_0^-\ \ket{\v^\pm_k}$
which have ghost number 3 and are annihilated by $b_0^-$
[\Ref{WZ},\Ref{ver}]. In the Batalin-Vilkovisky quantization these are
to be thought of as ``anti-tachyons'' [\Ref{ver}].

Now, commuting $Q_{BRST}$ in eq. \auno\ with the vertices and
propagators until it annihilates against the vacua, one gets a sum of
``canceled propagator contributions''. Each one has the form
$$
\sum_i\bra{V_{k_1}} V_{k_2} \Delta \cdots \Delta
V_{k_p}\ket{\Phi_i} \bra{\widetilde \Phi^i} V_{k_{p+1}}
\Delta cW_{J, m} \Delta \cdots V_{k_{n-1}} \ket{V_{k_n}} \ ,
\nfr{canp}
where $\ket{\widetilde\Phi^i}=b_0^-\ket{\Phi^i}$. By ghost number
counting, $\ket{\Phi_i}$ must have the physical ghost number 2, and,
therefore, $\ket{\widetilde\Phi^i}$ carries ghost number 3. Also,
the energy and momentum of $\ket{\Phi_i}$ are completely determined
because the correlation functions obey the sum rules eq. \sumrules.
It may happen that at this energy and momentum there is no physical
state. Then the usual ``canceled propagator argument'' applies
and we conclude that eq. \canp\ vanishes. In this theory
there is a class of cases, however, where there are physical states
contributing to the sum over $i$. Here we will only analyze the
situations where the energy and momentum that flow through the
``canceled propagator'' obey the tachyon dispersion relation.
Then the sum over $i$ collapses to one non-vanishing term involving
the tachyon states.

Note that eq. \canp\ can be interpreted as the
pinching of a sphere with $n$ punctures into
two spheres with $p$ and $n-p$ punctures respectively.
The amplitude corresponding to one of these spheres is perfectly
conventional, with each vertex operator carrying ghost number 2.
The other amplitude is unusual, since the current
operator $cW_{J, m}$ carries ghost number 1, while
$\ket{\widetilde\Phi^i}$ carries ghost number 3. In section 2.2 we will
show that this amplitude can be interpreted as a matrix
element of the charge operator ${\cal A}_{J, m}$.
For the correlation functions involving a current
we introduce the notation
$$\vev{\widetilde V_{-k} V_{k_1}\cdots V_{k_n} cW_{J, m}}=
\sum_P \bra{\widetilde V_{-k}} V_{k_1} \Delta cW_{J, m}
\Delta \cdots V_{k_{n-1}} \ket{V_{k_n}} \ .
\nfr{unusual}
The permutations $P$ involve all the vertex operators, except for
$\widetilde V$ which has to be kept as an out-state (in the
leftmost position). Since $\widetilde V$ and $cW$ are both
anti-commuting, this definition guarantees that all the
permutations contribute with the same sign. In general, in
dealing with correlators involving anti-commuting
vertex operators, their ordering becomes important, and
one may have to appeal to the operator formalism in order to
construct their proper definition.

The pinching of the sphere represented in eq. \canp\
is conformally equivalent to the situation where $p$ tachyon
vertex operators collide with $cW_{J, m}$, which is the language
used in ref. [\Ref{K}]. Using the {\scaps o.p.e.}, all the
colliding operators are replaced by insertion of a single
physical operator into the conventional amplitude. In eq. \canp\
this is the state $\ket{\Phi_i}$ that flows through the pinch.

Taking into account the sum over all permutations in eq. \auno,
the final form of the Ward identity is [\Ref{ver}]
$$0\ =\ \sum_{{\cal I}, {\cal L}}
\vev{V_{i_1} \dots V_{i_I} \Phi}
\vev{\widetilde\Phi\, V_{l_1}\cdots V_{l_L}\, cW_{J, m}}
\nfr{aund}
where the sum is over all possible partitions of the vertex operators
$V$ into two sets ${\cal I}$ and ${\cal L}$ with $I$ and $L$ elements
respectively.

Now, as we remarked earlier, there are cases when the kinematics
restricts $\Phi$ to be on the tachyon mass shell. Let us analyze
the energy and momentum conservation laws for the correlator
$$\vev{\widetilde\Phi\, V^+_{k_1}\cdots V^+_{k_n}\, cW_{J, m}}\ .
\nfr{ccorr}
If $\widetilde\Phi$ carries momentum $P$ and energy $E$, we obtain
$$\eqalign{&m+P+\sum_{i=1}^n k_i =0\cr
& J+E +\sum_{i=1}^n (-1+k_i) =-2\ .\cr}
\efr
For the number of particles
$n=J-m+1$, it follows that $E=-1+P$, which is the dispersion
relation of a positive chirality tachyon.
Remarkably, this applies for arbitrary momenta $k_i$, which is a
consequence of the two-dimensional kinematics!
Therefore, whenever $n=J-m+1$, eq. \aund\
receives a contribution from the canceled propagator with
${\widetilde \Phi}={\widetilde V^+_P}$, $\Phi= V^+_{-P}$.
Similarly, if we flip all the
chiralities, then the same situation arises for $n=J+m+1$. It is not
hard to see that these two cases are the only non-trivial amplitudes
of type \ccorr\ involving tachyons of generic (not discrete) momenta.
\chapter{Correlation functions from the Ward identities}
Now that we have reviewed the structure of the
Ward identities in the two-dimensional
closed string theory, and shown that they can give rise to non-trivial
constraints, we will write them out explicitly. In other words, we
will first explicitly calculate the unconventional
correlators of eq. \unusual. After substituting their values,
the Ward identities \aund\ become linear relations among the physical
tachyon correlators, which were derived in ref. [\Ref{K}].
We will focus on the tachyon correlators of type $(N, 1)$ and
$(1, N)$, which are the only ones that do not vanish for generic
momenta. These correlators are given by the multiple integrals
$$ \eqalignno{A_{N, 1}(k_1, \ldots, k_N)\ &=\
\vev{ V^-_{-k} (\infty)
V^+_{k_2}(1) \int T^+_{k_3} \dots \int T^+_{k_{N-1}}\,
V^+_{k_N}(0) }\ =\qquad\cr
=\  \prod_l &\int d^2 z_l \prod_i |z_i-1|^{4s_{i1}}
|z_i|^{4s_{i0}} \prod_{i<j} |z_i-z_j|^{4s_{ij}}&\nameali{gen}\cr }
$$
where $l$, $i$ and $j$ run from 2 to $N-1$, and
$$ s_{ij}=k_i k_j-\epsilon^+_i\epsilon^+_j=-1+k_i+k_j\ .
\efr
The integrals of eq. \gen\ are generalizations of the integrals
calculated in ref. [\Ref{Dots}]  by contour deformation techniques.
We will find that the integrals \gen\ can be calculated via simple
recursion relations which follow from the $w_\infty$ Ward identities.
This is a mathematical result which does not seem to be
well known.

To calculate the amplitude of eq. \unusual, we will observe
that it can be interpreted as the matrix
element of the charge operator ${\cal A}_{J,m}$ between a
$n$-tachyon state and a 1-tachyon state of the
form $\ket{\v^+_P}$. More explicitly, we will
show that~\note{A similar equation holds for the $T^-$ tachyons.}
$$\eqalignno{&
\vev{\widetilde V^+_{P}(\infty) \, c(w) W_{J,m}(w,\bar{w}) V^+_{k_1}
(0) \int T^+_{k_2} \dots \int T^+_{k_n}}\ = \qquad\qquad
&\nameali{asei}\cr &\qquad\qquad\qquad\qquad\qquad =\
\bivev{\v^+_{P}}{\,{\cal A}_{J,m} V^+_{k_1} (0)
\int T^+_{k_2} \dots \int T^+_{k_n}} \cr}
$$
where the charge operator acts on all tachyons to its right.

We will also present a different method of calculation of
of eq. \unusual, which relies directly on the Ward identities.
\section{Action of the $w_\infty$ charges on tachyons}
As a first step, we need to consider the kinematics of
the action of the charge ${\cal A}_{J,m}$ on $n$ tachyons
[\Ref{K},\Ref{WZ}]. Obviously, the analysis is the same as for the
correlators of eq. \unusual. Just from such kinematical considerations,
we conclude [\Ref{K}] that the charge
${\cal A}_{m+n-1,m}$ annihilates $l$ $T^+$ tachyons of generic
momenta, for $l<n$.
The first non-trivial action of this charge is on $n$
$T^+$ tachyons, producing only one $T^+$ tachyon [\Ref{K}]~:
$$
{\cal A}_{m+n-1,m} \, V^+_{k_1}(0) \int T^+_{k_2} \dots \int T^+_{k_n}
\ =\  F_{n,m}(k_1, \cdots ,k_n)\, V^+_k (0)
\nfr{action}
where $k= \sum k_i +m$.
A similar formula holds for ${\cal A}_{-m+n-1,m}$ acting on
$T^-$ tachyons.

The situation is more complex if one or more tachyons carry one of
the discrete momenta. For example,  let us consider the action of the
charges ${\cal A}_{m+n-1,m}$ on vertex operators $V^+_{s/2}$, with $s$
positive integers. For $n=1$ the action is correctly determined by
eq. \action. However, for $n>1$, the charge does not annihilate the
vertex operator. Instead, it produces a discrete state in the
``semi-relative" cohomology [\Ref{WZ}]. This discrete state
is not separately annihilated by $b_0$ and $\bar{b}_0$, but is
only annihilated by $b^-_0$. We will not analyze in detail such
action of the charges in this paper.

For a better understanding of the proof of eq. \asei\ we need to
review some of the results from refs. [\Ref{K},\Ref{WZ}].
Consider first the action of the charge ${\cal A}_{{\sssty 1\over
\sssty 2},-{\sssty 1\over\sssty 2}}$ on
two $T^+$ tachyons of generic momenta:
$$\eqalignno{
{\cal A}_{{\sssty 1\over\sssty 2},-{\sssty 1\over\sssty 2}}\,
V^+_{k_1} \int T^+_{k_2} =
&\left[\oint_\gamma {dz \over 2\pi i}\, W_{{\sssty 1\over\sssty 2},
-{\sssty 1\over\sssty 2}}
(z,\bar{z}) - \oint_\gamma {d\bar{z}\over 2\pi i} \, c(z)
\Psi_{{\sssty 3\over\sssty 2},-{\sssty 1\over\sssty 2}} (z)
\overline{X}_{{\sssty 1\over\sssty 2},-{\sssty 1\over\sssty 2}}
(\bar{z})\right] \cdot \cr
&\ \cdot\ V^+_{k_1}(0,0)\int d^2 w\, T^+_{k_2}(w,\bar{w})&\numali\cr}
$$
where the contour $\gamma$ encloses $0$. To get a
non-zero result from the action of the holomorphic part of the charge,
it is necessary that \noblackbox
$$
W_{{\sssty 1\over\sssty 2},-{\sssty 1\over\sssty 2}} (z,\bar{z})
V^+_{k_1}(0,0) \int d^2 w T^+_{k_2} (w,\bar{w}) = {1\over z}
F_{2,-\half}(k_1,k_2) V^+_{k_1+k_2
-{\sssty 1\over\sssty 2}}(0, 0)\ +\ \dots
\efr
The only contributions to the residue come from the
region where $z$ and $w$ approach each other and $0$. The
integral for $F_{2, -\half}(k_1,k_2)$ was calculated in ref.
[\Ref{K}] using the results of ref. [\Ref{Kawai}]
$$
F_{2,-\half}(k_1,k_2)\ =\ 2\pi(2k_1+2k_2 -1)
{\Gamma(1-2k_1) \over \Gamma(2k_1)}
{\Gamma(1-2k_2) \over \Gamma (2k_2) } {\Gamma(2k_1+2k_2-1) \over
\Gamma (2-2k_1-2k_2) }\ .
\efr
Also, in ref. [\Ref{WZ}] it was shown that the
anti-holomorphic part of the charge gives a contribution proportional
to $\bar{z}^s$ with $s> -1$. This is not singular enough to be relevant
and, therefore, the anti-holomorphic part of the charge does not
contribute in this case. One could anticipate this on general grounds
because this part of the charge carries (left,right) ghost numbers
equal to $(1, -1)$. The conservation of these ghost numbers forbids
the physical tachyon from appearing in the {\scaps o.p.e.}~.

In an analogous way one can compute the action of the charge ${\cal
A}_{m,m}$ on one $T^+_k$ tachyon (see ref. [\Ref{K}]). Using these
results and the $w_\infty$ algebra of charges, in ref. [\Ref{K}]
the action of the charge ${\cal A}_{m+n-1,m}$ on $n$ $T^+$ tachyons
was determined. The final result for eq. \action\ is \yesblackbox
$$
F_{n, m}(k_1, \dots, k_n)\ = \ 2\pi^{n-1} (n!)\, k
{\scriptstyle\Gamma(2k) \over \scriptstyle\Gamma(1-2k)}\,
\prod_{i=1}^n
{\scriptstyle\Gamma(1-2k_i) \over \scriptstyle\Gamma(2k_i)}\,
\nfr{explicit}
where $k=m+\sum_{i=1}^n k_i$ and $n\geq 1$.
A similar formula obviously holds for ${\cal A}_{-m+n-1,m}$
applied to $n$ generic $T^-$ vertex operators.
Note that an explicit evaluation
of this formula would require performing $n-1$ integrals, a very
difficult task, which is avoided here thanks
to the algebraic structure of the model.
\section{Calculation of the correlators involving a current}
In this section we will explicitly calculate the correlators of
eq. \unusual. We will do it in two different ways.
The first method involves the proof of
eq. \asei\ and then the use of eqs. \action\ and \explicit.
The second method is more direct: it relies on the application of
the Ward identities to the correlators involving a current.

Let us start by proving eq. \asei.
As we have seen in the previous section, the action of the charge is
completely characterized by the {\scaps o.p.e.}
of the holomorphic and anti-holomorphic currents with some number
of physical vertex operators. We will assume that the
following {\scaps o.p.e.} holds~:
$$
W_{J,m}(w, \bar{w})\,
V_{k_1}^+(0) \int T_{k_2}^+ \cdots \int
T_{k_n}^+  = {1\over w}\, F\,  V^+_k (0) +\ \dots\ .\ \
\nfr{Auno}

We will now show that the {\scaps r.h.s.} and the {\scaps l.h.s.}
of eq. \asei\ are equal. Consider the {\scaps r.h.s.}~:
$$\eqalignno{& \bivev{\v^+_{-k}}{\,{\cal A}_{J,m}
V^+_{k_1} (0) \int T^+_{k_2} \cdots \int T^+_{k_n}}\ =
&\numali\cr
&\quad =\ \bivev{\v^+_{-k}}{\left[\oint_\gamma
{dz\over 2\pi i}\, W_{J,m}(z,\bar{z}) - \oint_\gamma
{d\bar{z}\over 2\pi i}\, \Omega^{\bar z}_{J,m}(z,\bar{z})
\right] V^+_{k_1} (0) \int T^+_{k_2} \cdots \int
T^+_{k_n}} \cr}
$$
where the contour of integration $\gamma$ surrounds the point
$z=0$. From the previous section, we already
know, for example by ghost number conservation, that the
anti-holomorphic part of the charge, when acting on
generic tachyons, gives zero. Then, using eq. \Auno, one gets
$$\bivev{\v^+_{-k}}{\,{\cal A}_{J,m} V^+_{k_1} (0)
\int T^+_{k_2} \cdots \int T^+_{k_n}}\ =\ F\ .
\nfr{Asette}
Consider now the {\scaps l.h.s.} of eq. \asei~:
$$\eqalignno{&
\vev{\widetilde V^+_{-k}(\infty) c(w) W_{J,m}(w,\bar{w}) V^+_{k_1} (0)
\int T^+_{k_2} \cdots \int T^+_{k_n}}\ = &\nameali{ttemp}\cr
&\qquad\qquad\ =\ \bivev{\v^+_{-k}}{ \left[
\oint_\gamma {dz \over 2\pi i}\, z b(z)\ +\ \oint_\gamma {d{\bar z}
\over 2\pi i} \, \bar{z} \bar{b}(\bar{z}) \right] \ \cdot\cr
&\qquad\qquad\qquad\qquad\qquad \cdot\ c(w) W_{J,m}(w,\bar{w})
V^+_{k_1} (0) \int T^+_{k_2} \dots \int T^+_{k_n} } \cr}
$$
where we have written $b_0$ as $\oint_\gamma
{dz \over 2\pi i}\, z b(z)$, and $\gamma$ is the contour of
integration that wraps around the point $z=+\infty$.
Now we may deform the contour so that it surrounds all the other
points, i.e. $z=w$, $z=y_i$ and $z=0$.

As before, just by ghost number counting, the anti-holomorphic
contribution to eq. \ttemp\ vanishes. For the holomorphic integral,
we get
$$\eqalignno{ &\bivev{\v^+_{-k}}{ \oint_\gamma {dz
\over 2\pi i}\, z b(z) \, c(w) W_{J,m}(w,\bar{w}) V^+_{k_1} (0)
\int T^+_{k_2} \cdots \int T^+_{k_n} }\ =\qquad\qquad\quad\cr
&\qquad\qquad\qquad\qquad\qquad\ =\ \bivev{\v^+_{-k}}{
\oint_\gamma {dz \over 2\pi i}\, z { F\, V^+_k (0)\over (z-w)\, w }
}\ =  F\ . &\numali\cr}
$$
Thus, we have shown that eq. \asei\  holds. From eq. \action\ it
immediately follows that
$$\vev{\widetilde V^+_{-k} V^+_{k_1} V^+_{k_2} \cdots
V^+_{k_n} cW_{m+n-1, m}}\ =\  F_{n, m}(k_1, \dots , k_n)
\nfr{unusres}
where $k=m+\sum_{i=1}^n k_i$ and $F_{n,m}$ is given by eq. \explicit.

Now, we give another derivation of eq. \unusres, relying directly on
the existence of trivial gauge invariances. We start from the
Ward identities
$$\vev{\widetilde V^+_{-k}\{Q_{BRST}, cW_{J_1, m_1}\}
V^+_{k_1}\cdots V^+_{k_n} cW_{J_2, m_2}}\ =\ 0
\nfr{new}
where the kinematical constraints imply that
$$n=J_1+J_2-m_1-m_2+1\ ,\qquad\qquad k=m_1+m_2+\sum_{i=1}^n k_i\ .
\efr
We carefully define the correlator \new\ in the operator formalism,
summing over the permutations of all the operators except for
$\widetilde V^+_{-k}$, which is kept in the leftmost position.
Retaining all the cancelled propagator contributions, we find the
following recursion relation
$$\eqalignno{ 2\bigl ((J_1+1) m_2 & - (J_2+1) m_1\bigr)
\vev{\widetilde V^+_{-k}\, V^+_{k_1}\cdots V^+_{k_n}\,
cW_{J_1+J_2, m_1+m_2}}\ =~~&\nameali{newrec}\cr
=\ \ \ \ &\sum_{{\cal I}, {\cal L}} \vev{\widetilde V^+_{-k}\,
V^+_{i_1} \cdots V^+_{i_I}\, cW_{J_1, m_1} V^+_{k'}}
\vev{\widetilde V^+_{-k'}\, V^+_{l_1} \cdots V^+_{l_L}
\, cW_{J_2, m_2} } \cr
-\ &\sum_{{\cal R}, {\cal S}} \vev{\widetilde V^+_{-k}\,V^+_{r_1}
\cdots V^+_{r_R} \, cW_{J_2, m_2} V^+_{k'}}
\vev{\widetilde V^+_{-k'}\, V^+_{s_1} \cdots V^+_{s_S}
\, cW_{J_1, m_1} }\ .\cr}
$$
Let us explain the notation. $\sum_{{\cal I}, {\cal L}}$ means
the sum over all possible partitions of the $n$ tachyons into two sets
${\cal I}$ and ${\cal L}$ containing $I=J_1-m_1$ and
$L=J_2-m_2+1$ elements respectively. Similarly, $\sum_{{\cal R},
{\cal S}}$ means the sum over all possible partitions of the $n$
tachyons into two sets ${\cal R}$ and ${\cal S}$ containing
$R=J_2-m_2$ and $S=J_1-m_1+1$ elements respectively. In each term in
the sums, the momentum $k'$ is determined by the momentum conservation.
Finally, the $w_\infty$ structure constant on the left-hand side
arises from the 3-point function of two currents and one
``anti-current'' [\Ref{ver}].

The remarkable property of the recursion relations \newrec\
is that they only involve correlation functions of type \unusual,
i.e. those containing a single insertion of a current.
In order to solve these recursion relations, we need an input.
An example of the necessary input is all the three-point functions
and at least one four-point function. Here we may use the direct
calculations from ref. [\Ref{K}], which imply that
$$\eqalignno{
&\vev{\widetilde{V}^+_{-k_1-p}\, cW_{p,p} V^+_{k_1}} = 2(p+k_1)
\frac{\Gamma(2p+2k_1)}{\Gamma( 1-2p-2k_1)}
\frac{\Gamma( 1-2k_1)}{\Gamma( 2k_1)}  & \nameali{input}\cr
&\vev{\widetilde{V}^+_{-k_1-k_2+\half}\, c W_{{\sssty 1\over\sssty 2},
-{\sssty 1\over\sssty 2}}  V^+_{k_1} V^+_{k_2} } = 2\pi
(2k_1+2k_2-1) \frac{\Gamma( 2k_1+2k_2-1)}{\Gamma (2-2k_1 -2k_2)}
\frac{\Gamma( 1-2k_1)}{\Gamma( 2k_1)} \frac{\Gamma( 1-2k_2)}{\Gamma(
2k_2)}  \ .\cr}
$$

Let us illustrate the use of the relations \newrec\ by a simple
example, taking $(J_1, m_1)=(\half, -\half)$ and $(J_2, m_2)=(p,
p)$. Writing out eq. \newrec\ explicitly in this case, we find
$$\eqalignno{(4p+1) & \vev{\widetilde V^+_{-k}
V^+_{k_1} V^+_{k_2} cW_{p+\half, p-\half}}\ =&\numali\cr
&\qquad =\ \ \ \ \vev{\widetilde{V}^+_{-k}\, c W_{{\sssty
1\over\sssty 2}, -{\sssty 1\over\sssty 2}} V^+_{k_1} V^+_{k_2+p} }
\vev{\widetilde{V}^+_{-k_2-p}\, cW_{p,p} V^+_{k_2}} \cr
&\qquad\ \ \ +\  \vev{\widetilde{V}^+_{-k}\, c W_{{\sssty
1\over\sssty 2}, -{\sssty 1\over\sssty 2}} V^+_{k_2} V^+_{k_1+p} }
\vev{\widetilde{V}^+_{-k_1-p}\, cW_{p,p} V^+_{k_1}} \cr
&\qquad\ \ \ -\ \vev{\widetilde{V}^+_{-k}\, cW_{p,p} V^+_{k_1+k_2-
{\sssty 1\over\sssty 2}}} \vev{\widetilde{V}^+_{-k_1-k_2+
{\sssty 1\over\sssty 2}}\, c W_{{\sssty 1\over\sssty 2},
-{\sssty 1\over\sssty 2}}  V^+_{k_1} V^+_{k_2} } \cr}
$$
where, by kinematics, $k=k_1 +k_2 + p -\frac12$.
Substituting eq. \input, we immediately find
$$
\vev{\widetilde V^+_{-k}
V^+_{k_1} V^+_{k_2} cW_{p+\half, p-\half}}\ =\ 4\pi k\,
\frac{\Gamma( 2k)}{\Gamma (1-2k)}
\frac{\Gamma( 1-2k_1)}{\Gamma( 2k_1)}
\frac{\Gamma( 1-2k_2)}{\Gamma(2k_2)} \ .
\efr
Now, using eq. \newrec\ with $J_1=m_1+1$ and $J_2=m_2+1$,
we obtain the formula for correlation functions of $cW_{m+2, m}$,
and so on. This
recursive procedure is equivalent to that used in ref. [\Ref{K}]
in the language of $w_\infty$ charges acting on states. Here, following
refs. [\Ref{WZ},\Ref{ver}], we succeeded in reformulating
the recursion relations in terms of correlation functions.
Repeatedly using the recursion relations, we once again
arrive at eq. \unusres.
\section{Calculation of tachyon correlators via the Ward identities}
We can find all the correlation functions $A_{N,1}$, given by equation
\gen, from the following Ward identity
$$\vev{\{Q_{BRST}, cW_{-m, m}\}
V^+_{k_1}\cdots V^+_{k_N} V^-_{-\half}}=0
\nfr{newward}
where $N\geq 3$, and the kinematics fixes $m=1-\frac{N}2$.
Summing over all the non-vanishing canceled propagators, we find
$$\eqalignno{ 0\ =\ &\sum_{i=1}^N \vev{\widetilde V^+_{-k}
V^+_{k_{p(1)}} V^+_{k_{p(2)}} \cdots V^+_{k_{p(N-1)}}\,
cW_{-m, m}} \vev{V_k^+\, V^+_{k_i}
V^-_{-{\sssty 1\over\sssty 2}}} \cr
&\ +\ \vev{\widetilde V_{-r}^- \, V^-_{-{\sssty 1\over\sssty 2}}
cW_{-m, m}} \vev{V_{r}^-\, V^+_{k_1}
V^+_{k_2} \cdots V^+_{k_{N-1}}\, V^+_{k_N} }
&\nameali{Ward} \cr}
$$
where $p$ is the set of the first $N$ strictly positive integers except
for $i$, and $p(j)$ is the $j^{\rm th}$ element of the set.
The momenta of the intermediate
states, $r= \frac{1-N}2$ and $k= 1-\frac{N}2 + \sum_{j=1}^{N-1}
k_{p(j)}$, have been fixed
using the momentum and energy conservation laws. The form of the Ward
identity in eq. \Ward\ is identical to that in ref. [\Ref{K}], although
our starting point here was somewhat different. While ref. [\Ref{K}]
relied explicitly on the charge conservation, here we instead used the
technique of inserting a pure gauge state which is identically zero.

Now we use eq. \unusres, where $F_{n, m}$ is given by eq. \explicit, to
substitute the explicit expressions for the correlators involving the
current. Thus, we obtain a linear relation expressing the tachyon
$N+1$-point function in terms of the tachyon three-point function,
$$\eqalignno{
[(N-2)!]^2 (N-1) \, \cdot\, & A_{N,1}(k_1, \dots, k_N)
\ = &\nameali{all}\cr
&=\  \sum_{i=1}^N
F_{N-1, m} (k_{p(1)}, \dots , k_{p(N-1)} )\, \cdot\,
A_{2,1}(k, k_i )\ .\cr}
$$
Using the momentum conservation equation $2 \sum_{i=1}^N k_i =(N-1)$,
one gets
$$
F_{N-1, m} (k_{p(1)}, \dots , k_{p(N-1)} )\ =\
\pi^{N-2} (N-1)!\, (1-2k_i) \prod_{l=1}^N
\frac{\Gamma(1-2k_l)}{\Gamma(2k_l)} \ .
\efr
The three-point function $A_{2, 1}$ contains no integrations.
Therefore, it is independent of the momenta and is normalized as
$A_{2,1}=1$. Now eq. \all\ gives
$$
[(N-2)!]^2 (N-1) \, \cdot\, A_{N,1}(k_1, \dots, k_N)\ =\
\pi^{N-2} (N-1)!\,  \prod_{l=1}^N
\frac{\Gamma(1-2k_l)}{\Gamma(2k_l)}
\efr
from which we recover the correct answer [\Ref{GK},\Ref{kdf}]
$$
A_{N,1}(k_1, \dots, k_N)\ =\
{\pi^{N-2}\over (N-2)!} \prod_{l=1}^N
\frac{\Gamma(1-2k_l)}{\Gamma(2k_l)}\ .
\efr
If we change the normalization of the positive chirality
tachyons as in ref. [\Ref{K}], setting
$T^+_k = \frac{\Gamma(2k)}{\Gamma(1- 2k)} \exp (ikX+\epsilon_+\phi)$,
we get
$$
A_{N,1}(k_1, \dots, k_N)\ =\ {\pi^{N-2} \over (N-2)!}
\efr
in agreement with eq. (38) of ref. [\Ref{K}].
\chapter{Remarks on the open string case}
In this section we will sketch an extension of the Ward identities
to theories that include both closed and open strings.

As emphasized in ref. [\Ref{ver}],
the essential feature of the pure closed string case
is that a tube, pinched by a BRST charge, is replaced by insertion of
$\Phi_i$ on one side and $\widetilde\Phi^i=b_0^-\Phi^i$ on the other.
In the operator formalism, the remaining insertion of $b_0^-$ simply
follows from eq. \bzero. Thus, if $\Phi_i$ is annihilated by $b_0^-$,
so is $\widetilde\Phi^i$, i.e. both belong to the semi-relative
cohomology [\Ref{WZ}]. The resulting structure bears a strong
resemblance to the Batalin-Vilkovisky quantization scheme [\Ref{bv}],
where $\Phi_i$ and $\widetilde\Phi^i$ are to be thought of as field and
anti-field vertex operators [\Ref{ver}]. Since the sum of their
ghost numbers is 5, the corresponding field and anti-field couplings,
$\alpha^i$ and $\tilde\alpha_i$, carry opposite space-time
statistics. From the structure of the canceled propagator
contributions, it follows that the tree-level free energy $F$
satisfies the master equation [\Ref{ver}]
$$\sum_i{\partial F\over \partial \alpha^i}
{\partial F\over \partial \tilde\alpha_i}=0\ .
\nfr{meq}
This equation is a nice way to formalize the content of all
the tree-level closed string Ward identities.

Now we will show how to include the open string couplings in
the master equation. The structure of canceled
open string propagators is different from the closed string case,
and we need to take this into account. In open string theory,
the propagator is
$$\Delta_{open}={b_0\over L_0}\ .
\efr
Therefore, we get
$$\{Q_{BRST}, \Delta_{open}\}\ =\ {\bf 1}\ =\ \sum_i \ket{\phi_i}
\bra{\phi^i}\ .
\efr
Thus, unlike in the closed string case, there is no extra $b$ insertion.
The difference is obvious geometrically: when a tube is pinched, there
is a left-over modulus corresponding to the twist angle of the pinch;
when a strip is pinched, there are no left-over moduli.
Thus, a pinched strip is replaced by insertion of pairs of states that
are simply the conjugates of each other.
For instance, the state conjugate to $c\, t_k^\pm (0)\ket{0}$ is
$$c \partial c\, t^\pm_{-k}(0) \ket{0}
\efr
where the open string tachyon vertex operators are
$$\eqalign{& t^\pm_k\ =
\ e^{\half(ikX+\epsilon_\pm\phi)}\ ,\cr
&\epsilon_\pm=-1\pm k\ .\cr}
\efr
In general, if $\phi_i$ has ghost number $n$, then $\phi^i$ has
ghost number $3-n$, so that they have opposite statistic. Thus,
in the open string case, it is tempting to interpret $\phi^i$
as the anti-field vertex operator of $\phi_i$. The corresponding
coupling constants, $\beta^i$ and $\beta_i$, can then be thought of
as a space-time field and its anti-field.
Note that, if $b_0\ket{\phi_i}=0$, then
$b_0\ket{\phi^i}\neq 0$. Thus, in the open string Batalin-Vilkovisky
scheme we cannot impose the condition that all states are annihilated
by $b_0$, i.e. we have to include the whole absolute
cohomology.\note{Recall, for comparison, that in the closed string
sector the $b_0^-=0$ condition could be consistently imposed.}

Now that we have discussed the structure of the canceled open string
propagators, we can write down the generalization of the classical
master equation that should apply to the theory of coupled open and
closed strings,
$$\sum_i{\partial F\over \partial \alpha^i}
{\partial F\over \partial \tilde\alpha_i}+
\sum_j{\partial F\over \partial \beta^j}
{\partial F\over \partial \beta_j} \ =\ 0\ .
\nfr{nmeq}
We expect this equation to summarize all the Ward identities
that arise on a sphere and on a disc.

On a more detailed level, the Ward identities again follow from
the existence of trivial gauge invariances. Now, there are two
types of gauge invariances. In the open string sector, we have
$$\delta \Psi_{open}=[Q_{BRST}, \lambda_{open}]
\efr
where the physical ghost number of $\Psi_{open}$ is 1.
There are also the closed string gauge invariances of eq. \gauge.
Thus, there are two types of constraints on physical amplitudes of
open and closed strings. The first type follows from replacing
one of the open string vertex operators by
$$ [ Q_{BRST}, O_{J, m} ] =0\efr
where $O_{J, m}$ are the open string ground ring operators of ghost
number $0$ (they are analogous to the chiral ground ring of the
closed string).
The second type amounts to studying the effects of
the closed string symmetries. Here, as in the previous chapters,
we replace one of the closed  string vertex operators by
$\{Q_{BRST}, cW_{J, m}\}=0$.

In both cases we commute $Q_{BRST}$ through the propagators,
retaining the canceled propagator contributions whose sum has
to vanish. The analysis of kinematics, which is almost as simple as
in the pure closed string case, once again reveals many possibilities
of on-shell canceled propagators. The resulting
Ward identities are formally summarized in eq. \nmeq.
We expect these equations to provide stringent constraints on the
correlation functions, and to be a helpful tool in their calculations.
We will leave the detailed calculations for future
work.\note{Bershadsky and Kutasov have recently informed us that,
using somewhat different methods, they obtained constraints on the
correlation functions that, among other things, reproduce their
results in ref. [\Ref{BK}] (see also ref. [\Ref{tanii}]).}
\chapter{Discussion}
One may wonder, which elements of the structure of the Ward
identities are general, and which are special to the
two-dimensional model. As emphasized in refs. [\Ref{ver},\Ref{WZ}],
the structure of the canceled propagator contributions is in general
such that the pinch is replaced by insertions of field and
anti-field vertex operators. We have argued that the same conclusion
applies to models which include open strings. This simple analysis
implies a general connection between string Ward identities and
Batalin-Vilkovisky quantization.
However, the physical content of the Ward identities may be quite
trivial. In ref. [\Ref{K}], and in this paper, it was shown that this
is definitely not the case in the two-dimensional model. Here,
because of the presence of the discrete states that generate the
$w_\infty$ symmetry, and due to the special kinematics, the
Ward identities imply powerful recursion relations among the physical
correlation functions. These relations are, in fact, a valuable
practical tool that renders seemingly formidable calculations
easily doable.

In higher dimensions, because there is no infinite on-shell symmetry
such as $w_\infty$, the Ward identities are only
expected to establish the Poincar\'e invariance of the on-shell
amplitudes. One may hope, however, that the Ward identities have
more interesting manifestations off-shell [\Ref{WZ},\Ref{ver}].
This question needs to be explored further.
\acknowledgements
We thank C.~Nappi, A.~Polyakov, A.~Schwimmer, E.~Verlinde,
H.~Verlinde, E.~Witten, B.~Zwiebach and S.~Yankielowicz for helpful
discussions. The research of I.R.K. is supported in part by
DOE grant DE-AC02-76WRO3072, NSF Presidential Young Investigator
Award PHY-9157482, James S. McDonnell Foundation grant No. 91-48, and
an A.P. Sloan Foundation Research Fellowship. The research of A.P. is
supported by an I.N.F.N. fellowship.
\references
\beginref
\Rref{Kawai}{H.~Kawai, D.~Lewellen and S.H.~Tye, Nucl. Phys. {\bf B269}
(1986) 1.}
\Rref{GMil}{D.J.~Gross and N.~Miljkovic, Phys. Lett. {\bf 238B} (1990)
217; \newline
E.~Brezin, V.A.~Kazakov and Al.B.~Zamolodchikov, Nucl. Phys. {\bf B338}
(1990) 673;\newline
P.~Ginsparg and J.~Zinn-Justin, Phys. Lett. {\bf B240} (1990) 333;
\newline
G.~Parisi, Phys. Lett. {\bf B238} (1990) 209, 213.}
\Rref{GKleb}{D.J.~Gross and I.R.~Klebanov, Nucl. Phys.
{\bf B344} (1990) 375.}
\Rref{discrete}{J.~Goldstone, unpublished; V.G.~Kac, in {\sl Group
Theoretical Methods in Physics\/}, Lecture Notes in Physics, vol. 94,
Springer-Verlag, 1979.}
\Rref{GKN}{D.J.~Gross, I.R.~Klebanov and M.J.~Newman, Nucl. Phys.
{\bf B350} (1991) 621.}
\Rref{lz}{B.~Lian and G.~Zuckerman, Phys. Lett. {\bf 254B} (1991) 417,
Phys. Lett. {\bf 266B} (1991) 21.}
\Rref{aj}{J.~Avan and A.~Jevicki, Phys. Lett. {\bf 266B} (1991) 35,
Phys. Lett. {\bf 272B} (1991) 17, Mod. Phys. Lett. {\bf A7} (1992) 357;
\newline
D.~Minic, J.~Polchinski and Z.~Yang, Nucl. Phys. {\bf B369} (1992) 324;
\newline
G.~Moore and N.~Seiberg, ``{\sl From loops to fields in two
dimensional gravity\/}", preprint RU-91-29, YCTP-P19-91, July 1991;
\newline
S.~Das, A.~Dhar, G.~Mandal and S.~Wadia, ``{\sl Gauge theory formulation
of the $c=1$ matrix model: symmetries and discrete states\/}", preprint
IASSNS-HEP-91/52, October 1991, ``{\sl Bosonization of
non-relativistic fermions and W-infinity algebra\/}", preprint
IASSNS-HEP-91/72, November 1991, ``{\sl $W$-infinity Ward
identities and correlation functions in the $c=1$ matrix model\/}",
preprint IASSNS-HEP-91/79, December 1991.}
\Rref{Dots}{Vl.S.~Dotsenko and V.A.~Fateev,  Nucl. Phys. {\bf B251}
(1985) 291.}
\Rref{Sasha}{A.M.~Polyakov, Mod. Phys. Lett. {\bf A6} (1991) 635.}
\Rref{GK}{D.J.~Gross and I.R.~Klebanov, Nucl. Phys. {\bf B359} (1991)
3.}
\Rref{Goulian}{M.~Goulian and M.~Li, Phys. Rev. Lett. {\bf 66} (1991)
2051.}
\Rref{kdf}{P.~DiFrancesco and D.~Kutasov, Phys. Lett. {\bf 261B} (1991)
385, ``{\sl World sheet and space-time physics
in two-dimensional (super) string theory\/}", preprint PUPT-1276,
September 1991.}
\Rref{K}{I.R.~Klebanov, Mod. Phys. Lett. {\bf A7} (1992) 723.}
\Rref{W}{E.~Witten, ``{\sl Ground ring of two dimensional string
theory\/}", preprint IASSNS-HEP-91/51, August 1991.}
\Rref{KP}{I.R.~Klebanov and A.M.~Polyakov, Mod. Phys. Lett.
{\bf A6} (1991) 3273.}
\Rref{Polyakov}{A.M.~Polyakov, ``{\sl Singular states in 2d quantum
gravity\/}", preprint PUPT-1289, September 1991.}
\Rref{pert}{M.~Li, ``{\sl Correlators of special states in $c=1$
Liouville theory\/}", preprint UCSBTH-91-47, October 1991;\newline
J.~Barbon, ``{\sl Perturbing the ground ring of 2-d string theory\/}",
preprint CERN-TH6379-92, January 1992.}
\Rref{WZ}{E.~Witten and B.~Zwiebach, ``{\sl Algebraic structures and
differential geometry in 2d string theory\/}", preprint
IASSNS-HEP-92/4, MIT-CTP-2057, January 1992.}
\Rref{ver}{E.~Verlinde, ``{\sl The master equation of 2d string
theory\/}", preprint IASSNS-HEP-92/5, February 1992.}
\Rref{KMS}{D.~Kutasov, E.~Martinec and N.~Seiberg, ``{\sl Ground rings
and their modules in 2d gravity with $c\leq 1$ matter\/}", preprint
PUPT-1293, RU-91-49, November 1991.}
\Rref{Kachru}{S.~Kachru, ``{\sl Quantum rings and recursion relations
in 2d quantum gravity\/}", preprint PUPT-1305, January 1992.}
\Rref{divec}{P.~di~Vecchia, R.~Nakayama, J.L.~Petersen and S.~Sciuto,
Nucl. Phys. {\bf B282} (1987) 103;\newline
P.~di~Vecchia, M.~Frau, A.~Lerda and S.~Sciuto,
Nucl. Phys. {\bf B298} (1988) 526.}
\Rref{bv}{I.~Batalin and G.~Vilkovisky, Phys. Lett. {\bf 102B} (1981)
27,  Phys. Lett. {\bf 120B} (1983) 166.}
\Rref{tanii}{Y.~Tanii and S.~Yamaguchi, ``{\sl Two-dimensional quantum
gravity on a disk\/}", preprint STUPP-91-121, November 1991,
``{\sl Disk amplitudes in two-dimensional open string theories\/}",
preprint STUPP-92-128, March 1992.}
\Rref{GSW}{M.~Green, J.~Schwarz and E.~Witten, ``{\sl Superstring
Theory\/}", Vol. I and II, Cambridge University Press, Cambridge,
1987.}
\Rref{BK}{M.~Bershadsky and  D.~Kutasov, ``{\sl Open string theory in
1+1 dimensions\/}", preprint PUPT-1283, HUTP-91/A047, September 1991.}
\Rref{Liouv}{A.M.~Polyakov, Phys. Lett. {\bf 103B} (1981) 207.}
\Rref{ddk}{F.~David, Mod. Phys. Lett. {\bf A3} (1988)
1651;\newline
J.~Distler and H.~Kawai, Nucl. Phys. {\bf B321} (1989) 509.}
\endref
\ciao
